\documentclass[prb,aps,reprint,superscriptaddress,floatfix,showpacs,10pt]{revtex4-1}

\usepackage{graphicx,amsfonts,amssymb,amsmath,dsfont}
\usepackage{multirow}
\usepackage{times}
\usepackage{soul}
\usepackage[colorlinks,bookmarks=true,citecolor=blue,linkcolor=red,urlcolor=blue]{hyperref}

\usepackage{tikz}

\graphicspath{{Figs/}}

\newcommand{\ket} [1] {\vert #1 \rangle}

\def\ket#1{\vert#1\rangle}

\def\Longarrow{\protect\@lra}
\def\@lra{\relbar\joinrel\relbar\joinrel\relbar\joinrel%
          \relbar\joinrel\rightarrow}

\def\be{\begin{equation}}       \def\ee{\end{equation}}
\def\bea{\begin{eqnarray}}      \def\eea{\end{eqnarray}}
\def\bes{\begin{subequations}}  \def\ees{\end{subequations}}

\newcommand{\SCB}[0]{SrCu$_2$(BO$_3$)$_2$ }
\newcommand{\SCBn}[0]{SrCu$_2$(BO$_3$)$_2$}
\newcommand{\SSM}[0]{SSM }
\newcommand{\SSMn}[0]{SSM}
\newcommand{\AFM}[0]{AFM }
\newcommand{\AFMn}[0]{AFM}
\newcommand{\FM}[0]{FM }
\newcommand{\FMn}[0]{FM}
\newcommand{\EPP}[0]{EPP }
\newcommand{\EPPn}[0]{EPP}

\newcommand{\sv}[0]{\ensuremath{s_{\text{v}}} }

\newcommand{\svp}[0]{\ensuremath{s_{\text{v}}^{p}} }

\newcommand{\stt}[0]{\ensuremath{s_{\text{t}}} }

\newcommand{\stp}[0]{\ensuremath{s_{\text{t}}^{p}} }
\newcommand{\iPEPSn}[0]{iPEPS}
\newcommand{\iPEPS}[0]{iPEPS }

\newcommand{\SE}[0]{series expansions }

\begin{document}
\title{Exact plaquette singlet phases in an orthogonal-plaquette model}

\author{C.~Boos}
\email{carolin.boos@fau.de}
\affiliation{Institute for Theoretical Physics, FAU Erlangen-N\"urnberg, Germany}
%\affiliation{Institute of  Physics, Ecole Polytechnique F\'{e}d\'{e}rale de Lausanne (EPFL), CH 1015 Lausanne, Switzerland}

%\author{F.~Mila}
%%\email{frederic.mila@epfl.ch}
%\affiliation{Institute of Physics, Ecole Polytechnique F\'{e}d\'{e}rale de Lausanne (EPFL), CH 1015 Lausanne, Switzerland}

\author{K.~P.~Schmidt}
\email{kai.phillip.schmidt@fau.de}
\affiliation{Institute for Theoretical Physics, FAU Erlangen-N\"urnberg, Germany}

\date{\today}

\begin{abstract}
We introduce a quantum spin-$1/2$ model hosting two exact plaquette singlet ground states in extended parameter regimes.
There is an exact phase transition between both phases, at which the system has an extensive ground-state degeneracy.
Further, the model exhibits an extensive number of conserved quantities, which allow the prediction of additional phases.
We exploit this feature and explore the phase diagram in detail for two specific parameter regimes.
The model is based solely on Heisenberg interactions and seems sufficiently simple to be realized in quantum materials.
A general scheme to determine exact singlet product states is briefly discussed.
\end{abstract}

\maketitle

%%%%%%%%%%%%%%%%%%%%%%%
\section{Introduction}
\label{sec:Introduction}
%%%%%%%%%%%%%%%%%%%%%%%
Half a century ago, in a pioneering work, Majumdar and Gosh discovered a frustrated spin model hosting an analytically exact ground state~\cite{doi:10.1063/1.1664978}.
This opened a new pathway to exact results in frustrated spin systems, not only in one~\cite{doi:10.1063/1.1664978, PhysRevLett.59.799, PhysRevB.43.8644, MONTI1991197, PhysRevB.48.10653, doi:10.1143/JPSJ.64.2762, PhysRevB.56.R11380, PhysRevLett.78.3939, PhysRevLett.80.2709, Kolezhuk_1998, PhysRevB.69.094431}
% timers: Schmidt_2010
but also in two~\cite{SRIRAMSHASTRY19811069, PhysRevLett.93.217202, PhysRevB.77.014419, PhysRevLett.104.237201, doi:10.7566/JPSJ.86.054709} and three dimensions~\cite{Ueda_1999, Chen2002}.
%~\cite{SRIRAMSHASTRY19811069, PhysRevLett.59.799, MONTI1991197, PhysRevB.48.10653, PhysRevB.56.R11380, PhysRevLett.78.3939, PhysRevLett.80.2709, Kolezhuk_1998, PhysRevB.69.094431, PhysRevLett.93.217202, PhysRevB.77.014419, PhysRevLett.104.237201, Schmidt_2010, doi:10.7566/JPSJ.86.054709}.
%The Majumdar-Ghosh chain realizes coverings of singlets on dimers.
% as does the sawtooth-chain model~\cite{MONTI1991197}.
The specifically designed geometry and interactions enforcing exact states often lead to unconventional properties.
However, such models are not necessarily realized in materials.
One of the few examples, which directly describes a material, is the famous Shastry-Sutherland model (\SSMn)~\cite{SRIRAMSHASTRY19811069}.
It was invented out of purely theoretical interest, and only later found to capture the physics of the quasi two-dimensional quantum magnet \SCBn~\cite{PhysRevLett.82.3701, PhysRevLett.82.3168}.
This match led to a fast development in the field, since the exactness of the phase allowed a very precise theoretical understanding, and features like discretized magnetization plateaus could be studied~\cite{PhysRevLett.82.3701, PhysRevB.61.3417, doi:10.1143/JPSJ.69.1016, Kodama395, PhysRevB.93.241107}.
% 1999: miyahara
% 2000: doi:10.1143/JPSJ.69.1016
% 2002: kodama, FM science
% david: PhysRevB.93.241107
The \SSM is given by spins-$1/2$ coupled via Heisenberg interactions.
They are arranged as orthogonal dimers, which are connected by nearest-neighbor couplings forming a square lattice.
In the regime of strong dimers, the ground state is an exact product state of singlets on these dimers~\cite{SRIRAMSHASTRY19811069}.
For dominant nearest-neighbor couplings the antiferromagnetic ordered phase (\AFMn) occurs~\cite{SRIRAMSHASTRY19811069}.
In the intermediate regime a state of entangled singlets on empty (without an internal dimer bond) 4-spin plaquettes, the empty plaquette singlet phase (\EPPn), is stabilized\textcolor{red}{~\cite{PhysRevLett.84.4461}}.
Similar phases to the exact dimer singlet state in the \SSM are realized in all models referred to above,
and dimer singlets are by far the most commonly studied units in exact valence bond crystals~\cite{Miyaharaintrofrustmagn}.
Beyond dimer singlet states, models hosting exact dimer-tetramer~\cite{doi:10.7566/JPSJ.85.033705} and monomer-tetramer~\cite{PhysRevB.95.224415} states were {\color{red}}also introduced.
%Another class of exact states given by dimer singlets was described on lattices with relatively large unit cells, called quantum cages. Here, these states are extensively degenerate~\cite{PhysRevLett.104.237201}.
% sind das auch dimer singlets? singlet coverings, aber 2-fach entartet

The number of known frustrated quantum models with exact ground states is limited,
and to our knowledge, so far, no model hosting products of singlets on 4-spin plaquettes as an exact ground state has been proposed.
Such a model is desirable from a purely theoretical point of view, since it yields the possibility to gain a fundamental understanding of plaquette phases including excitations, correlations, and magnetizations.
The occurring properties are likely to reveal even more fascinating behavior than dimer singlet phases, due to the 16 states present on a decoupled 4-spin plaquette, in contrast to four states on a dimer.
This creates the possibility to tune between two distinct exact plaquette singlet phases via an exact phase transition.
Further, both plaquette singlets have different local properties, and allow a variety of triplon excitations~\cite{PhysRevB.62.5558, doi:10.1143/JPSJ.70.1369, PhysRevB.100.140413} as well as bound states.
% in plaquette singlet phases.
% question: cite for bound states?
%
Recently, plaquette phases in the neighborhood of a long-range ordered N\'eel phase were discussed in the context of a deconfined quantum critical point~\cite{sachdev2019}.
Experimentally, an entangled plaquette phase is realized in \SCB under external pressure~\cite{doi:10.1143/JPSJ.76.073710, Zayed_plaquette_16, doi:10.7566/JPSJ.87.033701, sandvik2019}, where further measurements are expected for an unambiguous identification of the phase, so a pristine theoretical understanding is in demand.
\begin{figure}[t!]
\begin{minipage}{\linewidth}
%(a)\\
\begin{center}
%  trim={<left> <lower> <right> <upper>}
\includegraphics[width=\linewidth, angle = 0, trim={3cm 21cm 2.3cm 2cm},clip]{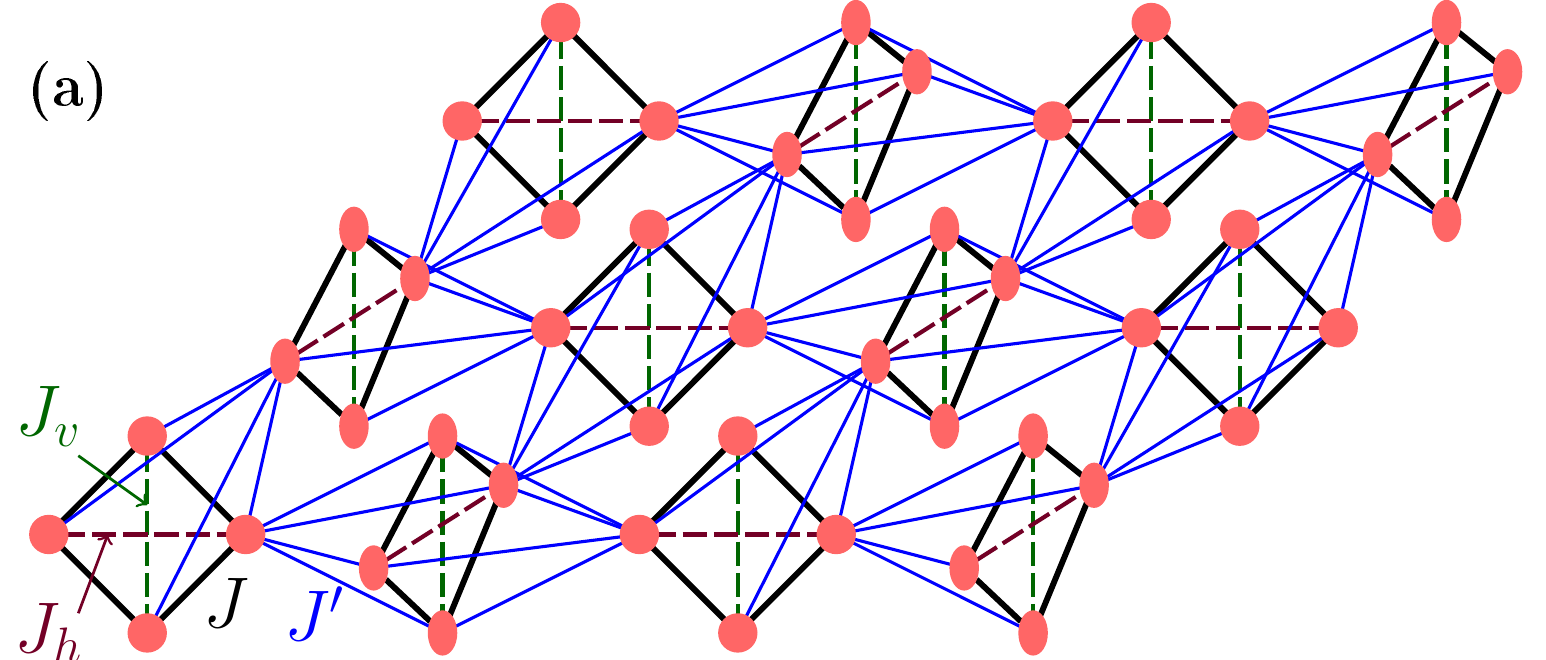}
\end{center}
\end{minipage}

\vspace{-0.1cm}

\begin{minipage}{\linewidth}
\begin{minipage}{0.65\linewidth}
\begin{flushleft}
%  trim={<left> <lower> <right> <upper>}
\begin{huge}
%\textbf{(b)}\\
\end{huge}
%  trim={<left> <lower> <right> <upper>}
\includegraphics[width=1\linewidth, angle = 0, trim={3cm 20.2cm 8cm 2cm},clip]{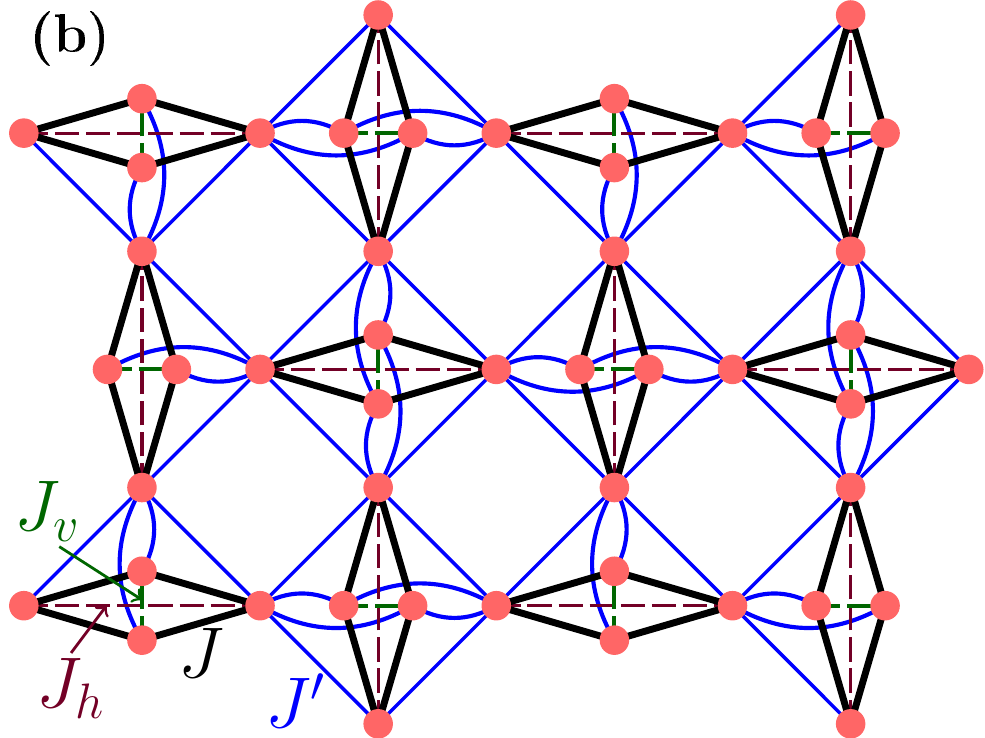}
\end{flushleft}
\end{minipage}%
\begin{minipage}{0.35\linewidth}
\begin{flushleft}
\begin{huge}
%\textbf{(c)}\\
\end{huge}
\end{flushleft}
\begin{flushright}
%  trim={<left> <lower> <right> <upper>}
\includegraphics[width=1\linewidth, angle = 0, trim={3cm 21.5cm 14cm 2cm},clip]{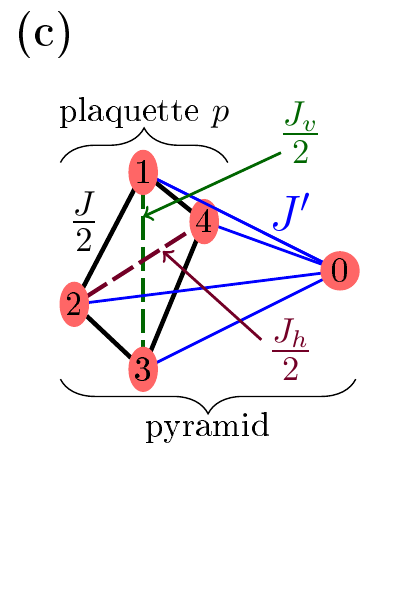}
\end{flushright}
\end{minipage}
\end{minipage}
\caption{(a): Orthogonal-plaquette model~\eqref{eq_hamilton} realizing exact plaquette product ground states with singlets on 4-spin $J$-plaquettes (solid black lines).
Red dots illustrate spins-$1/2$ interacting via Heisenberg interactions shown as lines.
The orthogonal $J_h$-bonds (dashed purple) together with the in-plane $J'$-bonds (solid blue) yield the \SSMn.
(b): Bird's eye view of the orthogonal-plaquette model~\eqref{eq_hamilton}. The $J'$-couplings present in the \SSM form a square lattice.
%The lattice arises from the \SSM given by the $J'$-square lattice (blue) and the orthogonal-dimers, $J_h$ (dashed purple), by extending these orthogonal-dimers to 4-spin plaquettes (4 red dots connected by black lines), compare with Fig.~\ref{fig_lattice}.
(c): A 5-spin pyramid with halved intra-plaquette couplings contains a 4-spin plaquette $p$.
The full model~\eqref{eq_hamilton} can be decomposed into a sum of such pyramids~\eqref{eq_hamilton_pyramids}.}
\label{fig_lattice}
\end{figure}

In this article, we introduce a quasi two-dimensional model hosting exact plaquette singlet ground states.
% plaquetten besonders interessant auch wegen SCB, DQCP
On top of that, the model offers an extensive number of conserved quantities, which enable some exact statements about other phases of the model.
%In this sense, it behaves somehow similar to orthogonal-dimer chains~\cite{0953-8984-10-16-015, PhysRevB.62.5558, doi:10.1143/JPSJ.70.1369}.
This allows the investigation of the phase diagram beyond the analytically determined area of the exact plaquette phases.
% several examples where theoreical models were found to be realized in materials this happened!
If the presented model is directly relevant for materials or can be simulated in experiments with artificial crystals remains open at this point.
However, this seems not implausible, since the model relies on nearest-neighbor Heisenberg exchanges only.
% based on the simplicity of the interactions.
%Historically, materials were synthesized, which realize models originally proposed out of purely theoretical motivation, just like \SCBn~\cite{SRIRAMSHASTRY19811069, PhysRevLett.82.3701, PhysRevLett.82.3168}.
%The specific model introduced here might be relevant in materials, since it relies on nearest-neighbor Heisenberg exchanges only.

%%%%%%%%%%%%%%%%%%%%%%%
\section{Orthogonal-plaquette model}
\label{sec:Orthogonal-plaquettemodel}
%%%%%%%%%%%%%%%%%%%%%%%
The quasi two-dimensional model consists of 4-spin plaquettes placed on a Shastry-Sutherland geometry.
Every dimer in the \SSM is replenished by a $45^\circ$-tilted plaquette $p$, which contains a vertical diagonal bond $J_v$ and the horizontal diagonal dimer bond $J_h$ from the \SSMn.
This orthogonal-plaquette model is illustrated in Fig.~\ref{fig_lattice}(a) and from a bird's-eye view in Fig.~\ref{fig_lattice}(b).
The Hamiltonian reads
\begin{equation}
\begin{aligned}
\label{eq_hamilton}
H= &J \sum_{\substack{p\\ \langle i,j \rangle}} \vec{S}_{p,i} \cdot \vec{S}_{p,j}
+ J_h \sum_{p} \vec{S}_{p,2} \cdot \vec{S}_{p,4}\\
+ &J_v \sum_{p} \vec{S}_{p,1} \cdot \vec{S}_{p,3} + J' \sum_{\substack{\left\langle p, p^\prime \right\rangle \\ \langle i,j \rangle }} \vec{S}_{p,i} \cdot \vec{S}_{p^\prime,j}\ ,
\end{aligned}
\end{equation}
where the spin operator $\vec{S}_{p,i}$ acts on spin $i$ of plaquette $p$. A plaquette $p$ is illustrated in Fig.~\ref{fig_lattice}(c).
The first sum runs over all spins on intra-plaquette nearest-neighbor bonds (black lines).
The second and third sum run over all plaquettes $p$ and the interactions address spins on horizontal diagonals (dashed purple lines) and on vertical diagonals (dashed green lines), respectively.
The fourth sum runs over all neighboring spins on different plaquettes $p$ and $p^\prime$ connected by inter-plaquette bonds (blue lines).
%The first sum runs over all nearest-neighbor (black lines), the second one over all horizontal diagonal (dashed purple lines), and the third one over all vertical diagonal (dashed green lines) intra-plaquette bonds.
%The fourth sum runs over all inter-plaquette bonds (blue lines) between neighboring spins on different plaquettes $p$ and $p^\prime$.
%
In the \SSMn, the exact singlets form on the $J_h$-diagonals.
In the orthogonal-plaquette model~\eqref{eq_hamilton} the exact singlets form on the 4-spin plaquettes of $J$-bonds, which are interconnected by $J'$-bonds.
The intra-plaquette couplings $J_h$ and $J_v$ introduce further triangles and lead to additional frustration.
Without these $J_h$- and $J_v$-bonds the model is already frustrated by triangles from an intra-plaquette bond (black lines) and two inter-plaquette bonds (blue lines).
The unit cell of the model contains eight spins of two orthogonally oriented plaquettes.
The model is invariant under rotations, $\mathcal{C}_4$, around center points between plaquettes.
Further, the reflection symmetries $\mathcal{R}_1$ and $\mathcal{R}_2$ over the perpendicular $J_h$-bonds as well as a full inversion over the $J_h$-plane hold.
The total spin on the vertical intra-plaquette $J_v$-diagonal, $\svp \in \{0,1\}$, is a good quantum number for \textit{every} plaquette $p$ individually, similar to the orthogonal-dimer chain~\cite{0953-8984-10-16-015}.
This can be traced back to the lattice structure, which does not have direct interactions between distinct diagonals.
In the following, $\sv$ denotes that all vertical diagonals are in the same state $\svp \ \forall \ p$ with the value $\sv \equiv \svp$.
Interestingly, if $\sv = 0$ the vertical dimer singlets are completely decoupled from all remaining sites, which form a Shastry-Sutherland lattice.
This can be seen for the according phases sketched in the phase diagrams in Fig.~\ref{fig_phasediagram_JveqJh} and Fig.~\ref{fig_phasediagram_Jh0}.

%%%%%%%%%%%%%%%%%%%%%%%
\subsection{Exact phases}
\label{subsec:Exactphases}
%%%%%%%%%%%%%%%%%%%%%%%
For $J'=0$ the orthogonal-plaquette model~\eqref{eq_hamilton} decouples into individual 4-spin plaquettes.
On such a plaquette $p$ the total spin of all four spins, $\stp \in \{0,1,2\}$, and the total spins on both diagonals, $\svp \in \{0,1\}$ and $s_{\text{h}}^p \in \{0,1\}$, give good quantum numbers.
An isolated plaquette $p$ exhibits two singlets $\stp=0$ with either two singlets ($\svp = s_{\text{h}}^p = 0$) or two triplets ($\svp = s_{\text{h}}^p = 1$) on the diagonals.
In order to prove the exact singlet plaquette ground states, we argue from two directions.
We start by showing that product states of plaquette singlets are eigenstates of the orthogonal-plaquette model~\eqref{eq_hamilton} and determine their ground-state energies.
Then, the lattice is separated into a sum of small units and a lower bound for the ground-state energy is derived.
%, which in some parameter regime is identical to the energies of the exact product states.

%%%%%%%%%%%%%%%%%%%%%%%%%%%%%%%%%%%%%%%%
i) All product states over plaquette singlets $\stp = 0 \ \forall \ p$ are exact eigenstates of the orthogonal-plaquette model~\eqref{eq_hamilton} since all inter-plaquette interactions can be written as \mbox{$\vec{S}_{p^\prime,0} \cdot (\vec{S}_{p, 1}+\vec{S}_{p,2}+\vec{S}_{p,3}+\vec{S}_{p,4})$} where the spin $\vec{S}_{p^\prime,0}$ belongs to one plaquette, and $\vec{S}_{p, 1}$ to $\vec{S}_{p, 4}$ form the neighboring plaquette (compare Fig.~\ref{fig_lattice}(c)).
The situation, where the total spin on all plaquettes is identical $\stp \ \forall \ p$ is denoted with $\stt \equiv \stp$.
Generically, $\stp$ is not a conserved quantity.
This only holds if $\stt = 0$ as for the product states of plaquette singlets.
The eigenstate where all plaquettes are in the same singlet $\stt = 0$ with $\sv = 0$ reads
\begin{align}
\label{eq_state_sd0_sp0}
\ket{\stt=0, \sv=0} &= \prod\limits_p \ket{ \stp=0, \svp=0}_p \ ,
\end{align}
whereas if $\stt = 0$ and $\sv = 1$ the eigenstate is given by
\begin{align}
\label{eq_state_sd1_sp0}
\ket{\stt=0, \sv=1} &= \prod\limits_p \ket{\stp=0, \svp=1}_p\ .
\end{align}
The eigenenergies per spin are
\begin{align}
\label{eq_eigenenergiessinglets_sd0}
&\epsilon^{\stt=0, \sv=0}=-3(J_h+J_v)/16 \quad \text{and}\\
\label{eq_eigenenergiessinglets_sd1}
&\epsilon^{\stt=0, \sv=1}=-J/2+(J_h+J_v)/16\ .
\end{align}
The corresponding eigenenergies of all other product states with combinations of plaquette singlets \mbox{$\stt = 0$} with distinct $s_{\text{v}}^p \neq s_{\text{v}}^{p^\prime}$ on different plaquettes $p$ and $p^\prime$ are only as low in energy as $\ket{\stt=0, \sv=0}$ and $\ket{\stt=0, \sv=1}$ where the energies of the latter two states cross.

ii) The Hamiltonian of the orthogonal-plaquette model~\eqref{eq_hamilton} can be decomposed into a sum over \mbox{5-spin} pyramids as shown in Fig.~\ref{fig_lattice}(c).
Let the spins of a pyramid be labeled by $\vec{S}_{p^\prime, 0}$ and $\vec{S}_{p, 1}$ to $\vec{S}_{p, 4}$.
Spins $\vec{S}_{p, 1}$ to $\vec{S}_{p, 4}$ are located on the 4-spin plaquette $p$ with nearest-neighbor interactions $J$, and diagonal $J_h$- and $J_v$-bonds between $\vec{S}_{p, 2}$ and $\vec{S}_{p, 4}$, and $\vec{S}_{p, 1}$ and $\vec{S}_{p, 3}$, respectively.
The additional spin $\vec{S}_{p^\prime, 0}$ from a neighboring plaquette $p^\prime$ interacts with all plaquette spins of $p$ with a coupling strength $J'$.
% Such a pyramid has the total spin of the plaquette $\stt$ and the total spin of the spins on the diagonal bond $\sv$ as good quantum numbers
The full Hamiltonian~\eqref{eq_hamilton} then reads
\def\modcdot{-0.025}
\begin{equation}
\label{eq_hamilton_pyramids}
\begin{aligned}
&H = \sum_{\text{pyramids}} \Bigg[ \frac{J}{2} \left(\vec{S}_{p, 1}+\vec{S}_{p, 3}\right) \hspace{\modcdot cm} \cdot \hspace{\modcdot cm} \left(\vec{S}_{p, 2}+\vec{S}_{p, 4}\right)\\
&+ \frac{J_h}{2} \vec{S}_{p, 2} \hspace{\modcdot cm} \cdot \hspace{\modcdot cm} \vec{S}_{p, 4} + \frac{J_v}{2} \vec{S}_{p, 1} \hspace{\modcdot cm} \cdot \hspace{\modcdot cm} \vec{S}_{p, 3}
+ J' \vec{S}_{p^\prime, 0}  \cdot \sum_{i=1}^4 \vec{S}_{p, i}\Bigg]\ .
\end{aligned}
\end{equation}
% ground states in general
A single pyramid containing the plaquette $p$ has two two-fold degenerate eigenstates with singlets on plaquettes, $\stp=0$, again distinguished by $\svp=0$ and $\svp=1$.
The degeneracy is manifested in the free spin $\vec{S}_{p^\prime, 0}$.
It occurs, since the $J'$-interactions do not contribute if $\stp=0$.
% Achtung: wir reden über eine Pyramide mit halbierten Bonds
The corresponding eigenenergies of a single pyramid are
\begin{equation}
\begin{aligned}
&E_0^{\stp = 0, \svp=0}=-3(J_h+J_v)/8 \quad \text{and}\\
&E_0^{\stp = 0, \svp=1}=-J+(J_h+J_v)/8\ .
\end{aligned}
\end{equation}
% some values
Depending on the parameter regime these plaquette singlets yield the degenerate ground states of the pyramid.
% evtl. todo: Grenzen inklusive ferro Fall angeben
For instance, the plaquette singlets with $\svp=1$ yield the ground states for
$J_h\leq J$, $J_v\leq J$, and $J'\leq J/2$, as well as for
$J_h=0$, $J_v \leq J$, and $J' \leq J-J_v/2$,
% $-J<J'<J/2$ at $J_v<J$
whereas the other singlets with $\svp=0$ are lowest in energy for
$J_h = J_v$, $J_v \geq J$, and $J'\leq J_v/2$.
Note, that ferromagnetic couplings are excluded, since the proof does not hold in this case.

\begin{figure}[t!]
\begin{center}
%\includegraphics[width=\linewidth]{plots/phasediagram_JhorizontalandJvertical.pdf}
%  trim={<left> <lower> <right> <upper>}
\includegraphics[width=\linewidth, trim={3.445cm 17.2cm 2.2cm 2.7cm},clip]{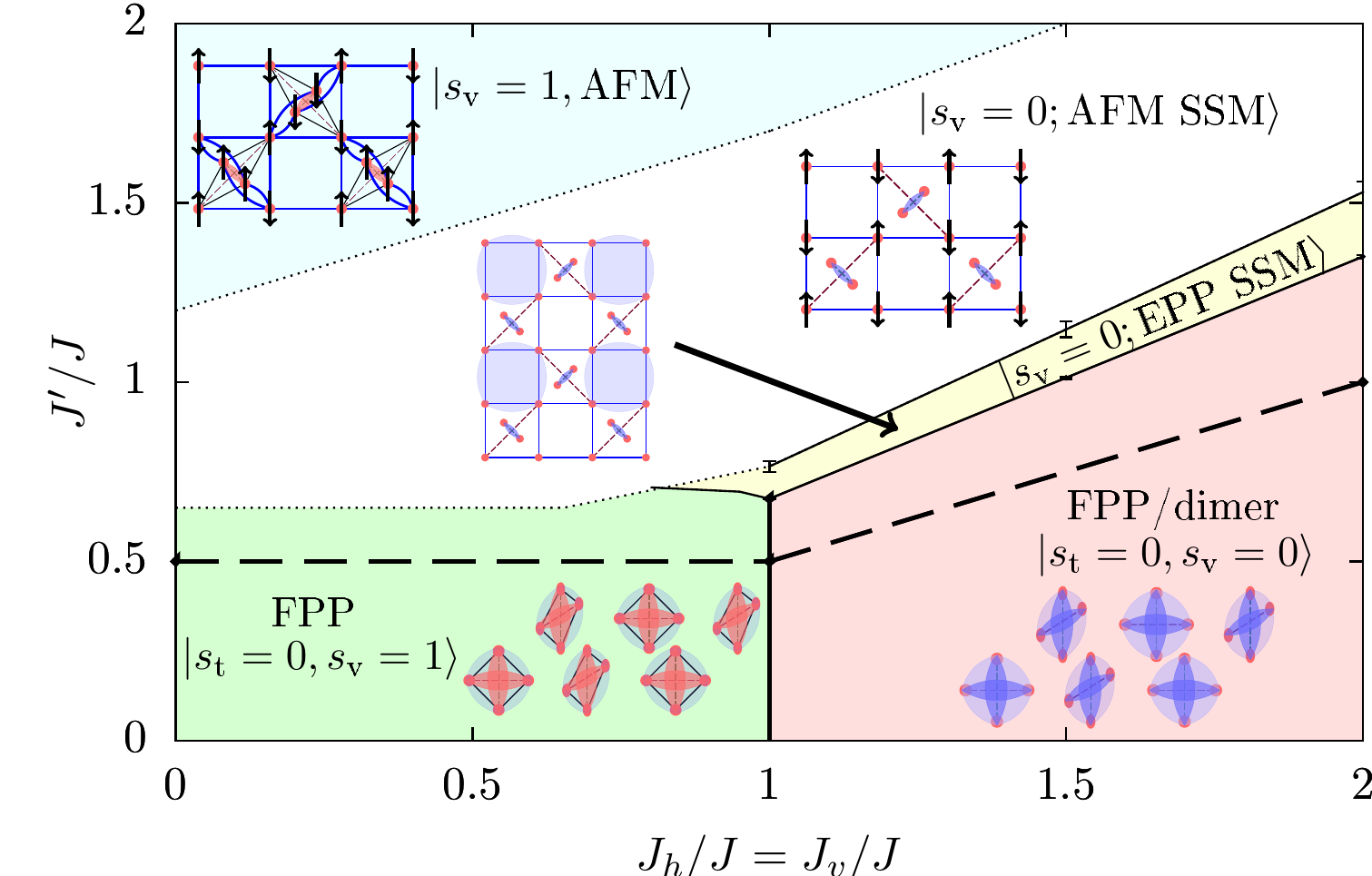}
\caption{Phase diagram of the orthogonal-plaquette model~\eqref{eq_hamilton} for $J_h=J_v$.
The exact plaquette singlet phases $\ket{\stt=0, \sv=1}$ (light-green region) and $\ket{\stt=0, \sv=0}$ (light-red region) are analytically proven to occur in the areas below the dashed black line.
The light-yellow and white areas yield the possible extension of an exact dimer singlet state $\sv = 0$ \mbox{($\svp = 0  \ \forall \ p$)} in a product with an \EPP or \AFM on the Shastry-Sutherland lattice, respectively.
All dotted lines are only sketched.
For the phase transition between \mbox{$\ket{\sv = 0; \text{\EPPn, SSM}}$} and $\ket{\sv = 0; \text{AFM, SSM}}$ as well as $\ket{\stt=0, \sv=0}$, numerical errors from iPEPS are also included~\cite{PhysRevB.87.115144}.
For the phase transition between \mbox{$\ket{\sv = 0; \text{\EPPn, SSM}}$} and $\ket{\stt=0, \sv=1}$, the uncertainties from the series expansions of the EPP ground-state energy are neglectable on this scale and are therefore not shown.
% The extension of the \AFM below $J_v/J<1$ as well as the upper bound in $J'/J$ are only sketched.
The \AFM with $\sv = 1$ (\mbox{$\svp=1 \ \forall \ p$}, light-blue region) is included, but its existence is only clear in the limit $J' \gg J, J_h, J_v$.
There might be other phases at intermediate coupling values.
%, in particular in the intermediate regime $J'\approx J_h \approx J$.
The given quantum numbers for all phases are exact.
% correlations
The phases are depicted such that couplings with vanishing contributions are not shown.
Entangled local units are shaded blue for singlets and red for triplets.
They are all exact apart from the plaquettes in $\ket{\sv = 0; \text{\EPPn, SSM}}$.}
\label{fig_phasediagram_JveqJh}
\end{center}
\end{figure}

%Note, that we exclude ferromagnetic couplings here since in that case it is not clear that a lower bound for the energies of the whole system, required in the next step, can be found.
% For $J_h=0$ a single pyramid realizes a two-fold degenerate ground state with a plaquette singlet \mbox{$\stt=0$} ($\sv=1$) for $-J<J'<J/2$ at $J_v<J$, and for $-2J+J_v<J'<J-J_v/2$ at $J_v>J$.
%
%
% on the full lattice
The lattice can be tiled with the plaquette singlet eigenstates of pyramids without ambiguity and the degeneracy due to $\vec{S}_{p^\prime, 0}$ is lifted.
The sum over all ground-state energies on 5-spin pyramids with plaquette singlets gives a lower bound for the ground-state energy of \eqref{eq_hamilton} for antiferromagnetic couplings.
This is based on the argument that joining pyramids cannot decrease the energy of the system:
On the one hand, two singlet plaquettes ''glued'' together do not change the energy per spin.
% On the other hand, connecting plaquette spins with different pyramids introduces further frustration, which can increase the energy of the system, or leave it unchanged if all contributions of $J'$-bonds vanish.
On the other hand, connecting spins on neighboring orthogonal plaquettes by $J^\prime$-bonds introduces further frustration, which increases the energy of the system.
The lower bounds (lb) for the eigenenergy per spin are
\begin{equation}
\begin{aligned}
\epsilon_{0,\text{lb}}^{\stt=0, \sv=0}&=  2 E_0^{\stp = 0, \svp=0}/4 \quad \text{and}\\
\epsilon_{0,\text{lb}}^{\stt=0, \sv=1}&= 2 E_0^{\stp = 0, \svp=1}/4\ .
\end{aligned}
\end{equation}
These energies are identical to the eigenenergies of the product over plaquette singlets in Eq.~\eqref{eq_eigenenergiessinglets_sd0} and Eq.~\eqref{eq_eigenenergiessinglets_sd1}.
Therefore, wherever a plaquette singlet state is the ground state on a single pyramid, the according product state over the lattice is the ground state of \eqref{eq_hamilton} when the interactions are antiferromagnetic.
% auf den pyramiden ist es erhalten, auch für ein triplet; ja? und sp=2 mit S0 anderer Farbe???
% example: Jh=Jv
For example, for $J_h=J_v$, the product state $\ket{\stt=0, \sv=1}$
is the exact ground state of the system for at least $0 \leq J' \leq J/2$ with $0 \leq J_v \leq J$, as illustrated by the light-green area below the dashed black line in the phase diagram in Fig.~\ref{fig_phasediagram_JveqJh}.
The product state $\ket{\stt=0, \sv=0}$ is the exact ground state for at least $0 \leq J'\leq J/2$ at $J_v \geq J$, which is captured by the light-red area below the dashed black line in the same phase diagram.
This case is directly related to the one in the \SSMn~\cite{SRIRAMSHASTRY19811069}.
Indeed, the exact state $\ket{\stt=0, \sv=0}$ can be seen as both, an exact plaquette singlet product state and an
exact dimer singlet product state, since all dimers are decoupled.
At the phase transition between the two exact singlet phases with $\sv = 0$ and $\sv = 1$, all products of plaquette singlets $\stt = 0$ with arbitrary combinations of $\svp = 0$ and $s_{\text{v}}^{p^\prime} = 1$, $p \neq p^\prime$, have the same eigenenergy.
Hence, the ground state is extensively degenerate.

%%%%%%%%%%%%%%%%%%%%%%%

%%%%%%%%%%%%%%%%%%%%%%%
\subsection{Phase diagrams}
\label{subsec:phasediagrams}
%%%%%%%%%%%%%%%%%%%%%%%
Whenever all spins on $J_v$-diagonals form singlets $\sv = 0$, these singlets are decoupled from the rest of the lattice.
This is illustrated from a bird's eye view for the phases with $\sv = 0$ in the phase diagrams in Fig.~\ref{fig_phasediagram_JveqJh} and Fig.~\ref{fig_phasediagram_Jh0}.
The vertical $J_v$-diagonals (dashed green bonds) are located in the center of the plaquettes (black bonds).
If $\sv = 0$ (singlets are shaded blue), all bonds between spins, which are connected with both spins of the $J_v$-bond vanish, i.e.~all plaquette \mbox{$J$-bonds} and half of the inter-plaquette $J'$-bonds.
This leads to the decoupling of the lattice into individual $J_v$-singlets and a \SSM of $J'$- and $J_h$-bonds,
%($J$ in the common \SSM notation)
which allows further insights on various phases beyond the exact plaquette singlet product states introduced above~\eqref{eq_state_sd0_sp0},~\eqref{eq_state_sd1_sp0}.
In the following, we discuss two special sets of parameters, namely $J_h=J_v$ and $J_h=0$.

%%%%%%%%%%%%%%%%%%%%%%%
\subsubsection{$J_h=J_v$}
\label{subsubsec:phasediagramsJveqJh}
%%%%%%%%%%%%%%%%%%%%%%%
For the symmetric model with $J_h=J_v$, we focus on antiferromagnetic couplings.
In this case, both exact phases are realized.
For weak inter-plaquette interactions $J'$ the exact plaquette singlet phase $\ket{\stt=0, \sv=1}$ is present, as illustrated by the light-green background color in the phase diagram in Fig.~\ref{fig_phasediagram_JveqJh}.
With increasing diagonal couplings an exact first-order phase transition towards the exact dimer singlet phase $\ket{\stt=0, \sv=0}$ (light-red region) occurs.
The transition line between the two is located at $J_h=J_v=J$ for at least $J'/J \leq 1/2$ as shown in Fig.~\ref{fig_phasediagram_JveqJh}.
Along this line the ground state is given by all product states of arbitrary combinations of plaquette singlets, $\svp=0$ and $s_{\text{v}}^{p^\prime}=1$ for $p\neq p^\prime$, and therefore has an extensive degeneracy.

In the dimer singlet phase, $\stt = 0$ and $\sv = 0$, the vertical $J_v$-dimers are decoupled from an independent \SSM of $J'$-bonds (solid blue) and $J_h$-bonds (dashed purple).
As long as $s_{\text{h}} = 0\ \forall \ p$, the $J'$-bonds do not contribute.
%%%%%%%%%%%%%%%%%%%%%%%%%%%%%%%%%%%%%%%%	EPP SSM
However, for increasing values of $J'$ beyond the dimer singlet phase, it is known that the \SSM realizes the entangled \EPPn~\cite{PhysRevB.87.115144}.
In the orthogonal-plaquette model~\eqref{eq_hamilton}, it occurs as a product state with additional dimer singlets
\begin{align}
\label{eq_sv0_epp}
\left |\sv=0;\ \text{EPP, SSM} \right\rangle = \left(\prod\limits_p \ket{\svp=0}_p \right) \left |\text{EPP}\right\rangle_{\text{SSM}}\ .
\end{align}
The phase transition between the dimer singlet phase and the \EPP in the \SSM is known from infinite projected entangled-pair states (\iPEPSn)~\cite{PhysRevB.87.115144},
which yields the transition line \mbox{$J'=0.675\pm 0.002 J_v$} for $J_v \geq J$.
The extension of the phase $\ket{\sv=0;\ \text{EPP, SSM}}$ is shown by the light-yellow background color in Fig.~\ref{fig_phasediagram_JveqJh}.
The errorbars from \iPEPS are included for three values of $J_h/J$.
For $J_v < J$, the dimer singlet phase does not occur, but the exact plaquette singlet phase $\ket{\stt=0, \sv=1}$.
In order to determine the phase transition between this phase and the \EPP in a product with $\sv=0$, we derived the energies of the \EPP up to order 9 in $J'/J$ and $J_h/J$.
For details on the series expansions, see the Appendix of Ref.~\onlinecite{PhysRevB.100.140413}.
We employ Pad\'e extrapolants with the order of the numerator $n$ and of the denominator $m$.
The average over three high-order extrapolants with $n=4, m=4$, $n=4, m=5$, and $n=5, m=4$ is considered as the most reliable resulting energy, and the uncertainty is estimated by the standard deviation.
For instance, for $J_h/J=J_v/J=0.95$, it is $J'/J |_c = 0.6946 \pm 0.0003$.
These small uncertainties are neglectable on the scale of the phase diagram in Fig.~\ref{fig_phasediagram_JveqJh}.

%%%%%%%%%%%%%%%%%%%%%%%%%%%%%%%%%%%%%%%%	AFM SSM
For larger $J'$-couplings in the \SSMn, the \EPP is replaced by the \AFMn.
Therefore, a product of an \AFM on the Shastry-Sutherland lattice with singlets on $J_v$-diagonals
\begin{align}
\label{eq_sv0_afm}
\left |\sv=0;\ \text{AFM, SSM}\right\rangle = \left(\prod\limits_p \ket{\sv=0}_p \right) \left |\text{AFM}\right\rangle_{\text{SSM}}
\end{align}
seems to be a good candidate phase, in particular at relatively weak plaquette couplings $J$.
Again, the transition to the \EPP is known from \iPEPSn~\cite{PhysRevB.87.115144}, $J'=0.765\pm 0.015 J_v$ for $J_v \geq J$, and the extension of the phase is illustrated by the white area in Fig.~\ref{fig_phasediagram_JveqJh}.

%%%%%%%%%%%%%%%%%%%%%%%%%%%%%%%%%%%%%%%% PURE AFM
In the limit $J=J_h=J_v=0$, the full model~\eqref{eq_hamilton} reduces to 4-spin plaquettes of $J'$-bonds.
The lattice is bipartite and one sublattice consists of all spins on parallel plaquettes.
In this regime also for finite $J$, $J_h$, and $J_v$, the model is expected to host an \AFMn~\cite{RevModPhys.63.1}, where all spins on one set of parallel plaquettes are effectively either up or down and inverse on the orthogonal plaquettes.
Thus, the total diagonal spin on every plaquette gives an exact triplet $\sv=1$, and we write the state as $\ket{\sv = 1, \text{AFM}}$.
Apart from quantum fluctuations, all spins point in the same direction on every plaquette and in the limit of decoupled $J$-plaquettes these states are connected to the quintuplet with $\stt=2$.
They form a square lattice of macro-spins (4-spin $J$-plaquettes).
This phase is indicated by a light-blue background in Fig.~\ref{fig_phasediagram_JveqJh}, even though we do not know any quantitative phase boundaries.

In the phase diagram in Fig.~\ref{fig_phasediagram_JveqJh} all phase boundaries shown as dashed lines are based on exact statements only.
All phase boundaries with dotted lines are only sketched.
All phase boundaries illustrated by solid lines are determined under the consideration of the discussed states [\eqref{eq_state_sd0_sp0}, \eqref{eq_state_sd1_sp0}, \eqref{eq_sv0_epp}, \eqref{eq_sv0_afm}] from exact statements together with numerical results, which include quantum fluctuations.
They are therefore expected to yield accurate results for the considered phases.
However, other phases with $\svp=1$ for some or all plaquettes $p$ can occur.
This seems in particular possible for small diagonal couplings $J_h$ and $J_v$.
For strong diagonal couplings these phases appear to be unlikely, and we expect the product phases with diagonal singlets~\eqref{eq_state_sd0_sp0},~\eqref{eq_sv0_epp}, and~\eqref{eq_sv0_afm} to be competitive energetically, since diagonal triplets induce further interactions and frustration.

%%%%%%%%%%%%%%%%%%%%%%%
\subsubsection{$J_h=0$}
\label{subsubsec:phasediagramsJhzero}
\begin{figure}[t!]
\begin{center}
%\includegraphics[width=\linewidth]{plots/phasediagram_larger.pdf}
%  trim={<left> <lower> <right> <upper>}
\includegraphics[width=\linewidth, trim={3.32cm 16.1cm 2.2cm 2.6cm},clip]{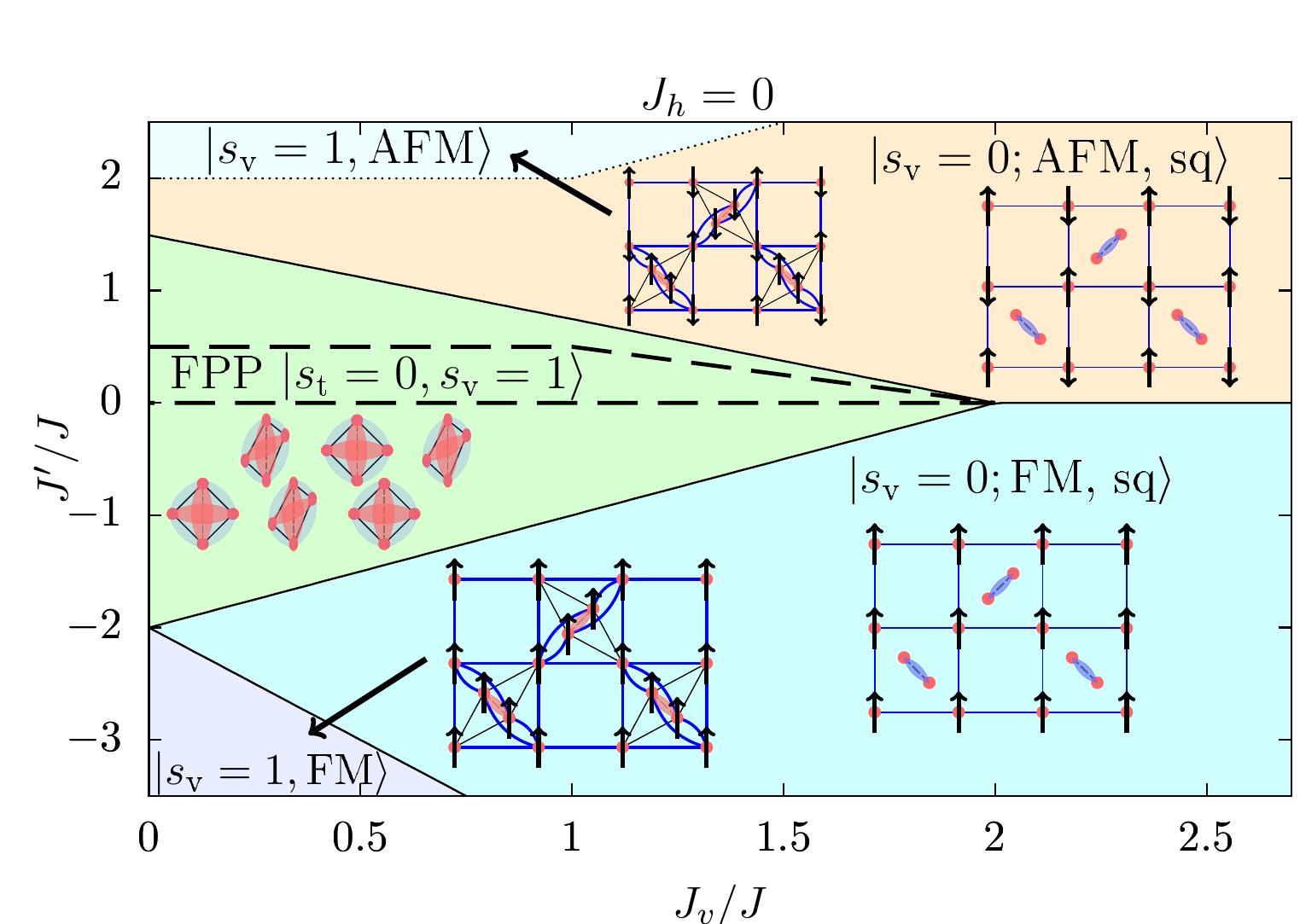}
\caption{Phase diagram of the orthogonal-plaquette model~\eqref{eq_hamilton} with $J_h=0$.
The exact plaquette singlet phase $\ket{\stt=0, \sv=1}$ (light-green) is analytically proven to occur in the area surrounded by the dashed black line.
The light-orange and light-cyan areas yield the extension of products from dimer singlets $\sv = 0$ with an \AFM and \FMn, respectively.
A pure \FM is shown in light-gray.
There might be other phases in these areas.
The \AFM with $\sv = 1$ ($\svp=1 \ \forall \ p$, light-blue) is sketched, and is only clear in the limit $J' \gg J, J_v$.
Quantum numbers and correlations can be understood as in Fig.~\ref{fig_phasediagram_JveqJh}.}
\label{fig_phasediagram_Jh0}
\end{center}
\end{figure}

%%%%%%%%%%%%%%%%%%%%%%%
Next, we discuss the orthogonal-plaquette model~\eqref{eq_hamilton} with vanishing horizontal diagonal couplings $J_h=0$ for ferromagnetic and antiferromagnetic inter-plaquette interactions $J'$.
Again, for small diagonal couplings, the exact plaquette singlet phase $\ket{\stt=0, \sv=1}$ is realized.
The smallest possible extension of this phase is illustrated by the dashed black line in the phase diagram in Fig.~\ref{fig_phasediagram_Jh0}.
For $J_h=0$ the exact dimer singlet phase~\eqref{eq_state_sd0_sp0} does not occur, unless $J'=0$, as explained in the following.

With $\sv = 0$ the system decouples into \mbox{$J_v$-singlets} and a separate square (sq) lattice of $J'$-bonds.
This is clear, since coming from the \SSM only the diagonal $J_h$-couplings are removed.
%
%%%%%%%%%%%%%%%%%%%%%%%%%%%%%%%%%%%%%%%% S_P=0 S_D=0
If all plaquettes are in a singlet state $\stt=0$ with $\sv=0$, the $J'$-couplings do not contribute either.
% s_p=1
In contrast, if the plaquettes are not in a singlet state, the $J'$-square lattice is present and leads to a reduction of the energy.
Hence, the exact dimer singlet phase $\ket{\stt=0, \sv=0}$ does not occur for $|J'|>0$.
% ever!
%
%%%%%%%%%%%%%%%%%%%%%%%%%%%%%%%%%%%%%%%% ANTIFERROMAGNET + sd=0
For antiferromagnetic inter-plaquette couplings, $J'>0$, an \AFM occurs on the square lattice~\cite{RevModPhys.63.1}.
The ground-state energy can be taken from \SE \mbox{$\epsilon_0^{\text{\AFMn, sq}}=(-0.6696 \pm 0.0003)J'$}~\cite{PhysRevB.39.9760}.
The product state from the \AFM on the square lattice and $J_v$-singlets is written as
\begin{align}
\label{eq_state_sv0_afm_sq}
\left|\sv=0;\ \text{\AFMn, sq}\right\rangle = \left(\prod\limits_p \left|\sv=0\right\rangle_p \right) \ket{\text{\AFMn}}_{\text{sq}}\ .
\end{align}
The eigenenergy per spin is
\begin{align}
\epsilon^{\sv =0; \text{\AFMn, sq}}=(-3J_v/16 + \epsilon_0^{\text{\AFMn, sq}})/2\ .
\end{align}
%%%%%%%%%%%%%%%%%%%%%%%%%%%%%%%%%%%%%%%% FERROMAGNET + sd=0
For $J'<0$ the square lattice exhibits a ferromagnetically ordered  phase (\FMn) with $\epsilon_0^{\text{\FMn, sq}}=J'/2$.
The corresponding product state in the orthogonal-plaquette model~\eqref{eq_hamilton} reads
\begin{align}
\label{eq_state_sv0_fm_sq}
\left|\sv=0;\ \text{\FMn, sq}\right\rangle = \left(\prod\limits_p \ket{\sv=0}_p \right) \left|\text{\FMn}\right\rangle_{\text{sq}}
\end{align}
and has the eigenenergy per spin
\begin{align}
\epsilon^{\sv=0;\text{\FMn, sq}} = -3J_v/16 + J'/4\ .
\end{align}
%%%%%
% motivate the connection of a local plaquette with AFM und FM sd=0
%The \AFM and \FM on a square lattice, $s_{\text{d}}=0$ phases in the limit of decoupled plaquettes $J'=0$ are both connected to the same triplet state, with a triplet on the horizontal diagonal ($\vec{S}_2+\vec{S}_4$).
In the limit of decoupled plaquettes, $J'=0$, both product states of dimer singlets and magnetic order on a square lattice, $\ket{\sv=0;\ \text{\AFMn, sq}}$ and $\ket{\sv=0;\ \text{\FMn, sq}}$, are connected to the same plaquette triplet state with $s_{\text{h}}^p = 1\ \forall \ p$.
They are, however, distinguished by the total spin in $z$-direction \mbox{$s_z^p\neq s_z^{p^\prime}$} on distinct plaquettes $p$ and $p^\prime$.
In the \FM case, all spins on $J_h$-bonds point in the same direction, and a single plaquette state tiles the lattice.
In contrast, in the \AFM case the states on perpendicular plaquettes $p$ and $p^\prime$ have to be chosen from different sectors, for instance $s_z^p= 2$ and $s_z^{p^\prime}= -2$.

%%%%%%%%%%%%%%%%%%%%%%%
%%%%%%%%%%%%%%%%%%%%%%%

%%%%%%%%%%%%%%%%%%%%%%%%%%%%%%%%%%%%%%%% PURE FM
Eventually, for strong ferromagnetic couplings $J'<0$ with $|J'|\gg J, J_v$ the ferromagnetic phase is expected with the ground-state energy per spin
\begin{align}
\epsilon_0^{\text{\FM}}=J/4+J_v/16+3J'/8\ .
\end{align}
% comparison
The comparison of all derived eigenenergies yields the phase diagram as indicated by the background colors in Fig.~\ref{fig_phasediagram_Jh0}.
Again, apart from the regime where the exact plaquette phase was proven, the occurrence of other phases [other than \eqref{eq_state_sd0_sp0}, \eqref{eq_state_sd1_sp0}, \eqref{eq_state_sv0_afm_sq}, \eqref{eq_state_sv0_fm_sq}] can not be excluded, but seems implausible for strong $J_v$-couplings.
For the considered phases, quantum fluctuations either persist only on local units [\eqref{eq_state_sd0_sp0}, \eqref{eq_state_sd1_sp0}], vanish [\eqref{eq_state_sv0_fm_sq}], or are taken into account numerically [\eqref{eq_state_sv0_afm_sq}].
The error on the energy of $\left|\sv=0;\ \text{\AFMn, sq}\right\rangle$ from series expansions is neglectable on the scale of the phase diagram.
The extension of the antiferromagnetic phase $\ket{\sv = 1, \text{AFM}}$ is sketched and only clear in the limit $J' \gg J_v, J$.

%%%%%%%%%%%%%%%%%%%%%%%
\section{Conclusion and outlook}
\label{sec:outlook}
%%%%%%%%%%%%%%%%%%%%%%%
We introduced a quasi two-dimensional orthogonal-plaquette model with four Heisenberg couplings.
All products of plaquette singlets were shown to yield exact eigenstates.
The product states of a single type of plaquette singlet $\ket{\stt=0, \sv=0}$ and $\ket{\stt=0, \sv=1}$ were proven to be the ground states, wherever the building block of the lattice, a 5-spin pyramid [Fig.~\ref{fig_lattice}(c)], has such a plaquette singlet ground state.
These states are most likely more extended than on a single pyramid, as it was also found for the dimer singlet phase in the \SSMn~\cite{SRIRAMSHASTRY19811069, PhysRevB.87.115144}.
All products of plaquette singlets constitute the extensively degenerate ground-state manifold at phase transitions between the two plaquette singlet ground states $\ket{\stt=0, \sv=1}$ and $\ket{\stt=0, \sv=0}$.
Further, there is an extensive number of conserved quantities given by the total spin on the $J_v$-diagonal on every plaquette.
We exploited this property and studied the phase diagram beyond the exact plaquette singlet phases.
For $J_h=J_v$ and antiferromagnetic couplings, products between \mbox{$J_v$-singlets} $\sv = 0$ and phases of the \SSMn, i.e.~an entangled \EPP and an \AFMn, seem most likely.
If $J_h=0$, the Shastry-Sutherland lattice reduces to a square lattice.
In this case, products of $J_v$-singlets with \FM or \AFM are realized.
In the large intra-plaquette coupling limit, $J'\gg J, J_h, J_v$, an \AFM with $\sv=1$ occurs that is connected to quintuplet states $\stt =2$ on individual plaquettes.
% The purely \FM state is also present in the model. trivial!
The search for additional phases is left for future investigations.
To this end, apart from exact diagonalizations, series expansions, and iPEPS, for not too large values of $J'$ Quantum Monte-Carlo simulations should be possible, due to the exact ground states.
This route was recently taken for the \SSM in a dimer singlet basis~\cite{PhysRevB.98.174432}.

From the exact properties of the orthogonal-plaquette model, some further statements can be made.
% excitations
In the exact plaquette singlet phase $\ket{\stt=0, \sv=1}$ two triplon modes occur as expected from the limit of disconnected 4-site plaquettes.
One of these modes is localized since it has $\sv^{p^\prime}=0$ on a single plaquette $p^\prime$, whereas all other plaquettes have $\svp=1 \ \forall \ p \backslash p^\prime $.
The associated one-triplon dispersion is completely flat.
The other mode has $\svp=1$ and is expected to be dispersive.
For small $J_v$- and $J_h$-couplings, the $\svp=1$ excitation should host the gap, whereas for larger couplings $J_v$ and $J_h$ it is the excitation with $\svp=0$.
Such a behavior is known from orthogonal-dimer chains~\cite{PhysRevB.62.5558}. 
For $J_h=0$, it is clear that the exact dimer singlet phase in an \AFM background, $\ket{\sv=0;\ \text{\AFMn, sq}}$, has a gapless magnon continuum.
The singlets on $J_v$-dimers yield dispersionless triplon excitations with $\svp=1$.
However, for $\svp=1$ the couplings between dimers and the square lattice do not vanish.
For a better and quantitative understanding further studies are required.

We note that the construction of models hosting exact product states can be generalized:
Instead of 4-spin plaquettes, other units realizing a total singlet ground state can be taken.
For every pair of interacting spins from two different units on the lattice, at least one of the spins must couple to all spins of this neighboring unit homogeneously.
The lattice must be decomposable into a sum over enlarged units (3-spin triangles for 2-spin dimers, 5-spin pyramids for orthogonal 4-spin plaquettes).
Then, if the connection of these enlarged units to the full lattice overall increases the frustration, the product state of singlets on the units yields the ground state of the full model in at least the parameter range, where the singlets determine the ground state of the enlarged unit.
For instance, the orthogonal structure of the \SSM leads to exact valence bond crystals for all plaquettes with local singlets, not only dimers and 4-spin plaquettes, but also hexagons, octagons, or larger shapes.
These have an even richer local structure, but seem less likely to be realized in materials.
% However, to this date we did not find a three-dimensional extension of the model, in which the exact plaquette state is protected.
% experiment
In contrast, the quasi two-dimensional orthogonal plaquette model, especially in the symmetric case with $J_h=J_v$, seems simple enough that an experimental realization could be possible, either in a material or an artificial system.
This would yield a good framework for the combined understanding of plaquette phases in theory and experiments.

%%%%%%%%%%%%%%%%%%%%%%%%%%%%%%%%%%%%%%%%%%%%%%%%%%%%%%%%%%%%%%%%%%%%%%%%%%%%%%%%
% Acknowledgement
%%%%%%%%%%%%%%%%%%%%%%%%%%%%%%%%%%%%%%%%%%%%%%%%%%%%%%%%%%%%%%%%%%%%%%%%%%%%%%%%
\section*{Acknowledgement}

We thank Fr\'ed\'eric Mila for fruitful discussions.
We acknowledge financial support by the German Science Foundation (DFG) through the Engineering of Advanced Materials Cluster of Excellence (EAM) at the Friedrich-Alexander University Erlangen-N\"urnberg (FAU).
%, and by the Swiss National Science Foundation.

%\appendix

%%%%%%%%%%%%%%%%%%%%%%%
%\section{CSL for $\Phi\neq \pi$}
%\label{sec:CSLnonpiflux}
%%%%%%%%%%%%%%%%%%%%%%%

\bibliography{exact_plaquettes}

\end{document}